\documentclass[a4paper]{article}
\usepackage{graphicx}

\begin{document}

\begin{center}

{\Large \bf The effect of uu diquark suppression in proton splitting in
            Monte Carlo event generators}
 
\end{center}

\begin{center}
{V. Uzhinsky\footnote{CERN, Geneva, Switzerland}$^,$\footnote{Laboratory of Information Technologies, JINR, Dubna, Russia},
 A. Galoyan\footnote{Veksler and Baldin Laboratory of High Energy Physics, JINR, Dubna, Russia}}
\end{center}

\begin{center}
\begin{minipage}{14cm}
Monte Carlo event generators assume that protons split into (uu)-diquarks and
d-quarks with a probability of 1/3 in strong interactions. It is shown in this paper
that using a value of 1/6 for the probability allows one to describe at
a semi-quantitative level the NA49 Collaboration data for $p+p\rightarrow p+X$ reactions
at 158 GeV/c. The Fritiof (FTF) model of Geant4 was used to simulate the reactions.
The reduced weight of the (uu)-diquarks in protons is expected in the instanton model.
\end{minipage}
\end{center}

\vspace{5mm}\noindent PACS: 24.10.Lx, 13.85.-t,13.85.Ni, 14.20.-c \vspace{5mm}

Most of the Monte Carlo event generators of multi-particle production assume that
nucleons split into diquarks and quarks in strong interactions. In particular,
protons split into (ud)-diquarks and u-quarks with a probability of 2/3, and into
(uu)-diquarks and d-quarks with a probability of 1/3. At the same time, there are
various physical signatures that the probabilities can be different \cite{1}. For
example, it was assumed in the papers \cite{2}, as in many other papers, that
the (ud)--u configuration completely dominates in the proton wave function. This was
motivated be the instanton model \cite{3} of the QCD vacuum. According to that model,
quark-quark interactions are flavor-dependent: they are non-zero only if quarks
have different flavors. Thus, (uu)-diquarks must be suppressed in protons \cite{4}.
The true weight of the (uu)--d configuration can be estimated using the NA49 Collaboration
data \cite{5}.

The NA49 Collaboration presented high precision data on particle production in
$pp$ interactions at 158 GeV/c including $x_F$, $p_T$ and rapidity distributions of
various particles ($p$, $n$, $\pi^\pm$, $K^\pm$, $\bar p$). As shown in \cite{6,7}, Monte
Carlo event generators based on the Fritiof model \cite{8,9} cannot satisfactorily
describe the data. The most dramatic situation takes place with a description of
the proton spectra. A typical prediction for the $x_F$-spectrum is shown in Figure~1
and is presented by the solid thin line.
\begin{figure}[cbth]
\begin{center}
\begin{minipage}{14cm}
\includegraphics[width=140mm,height=70mm,clip]{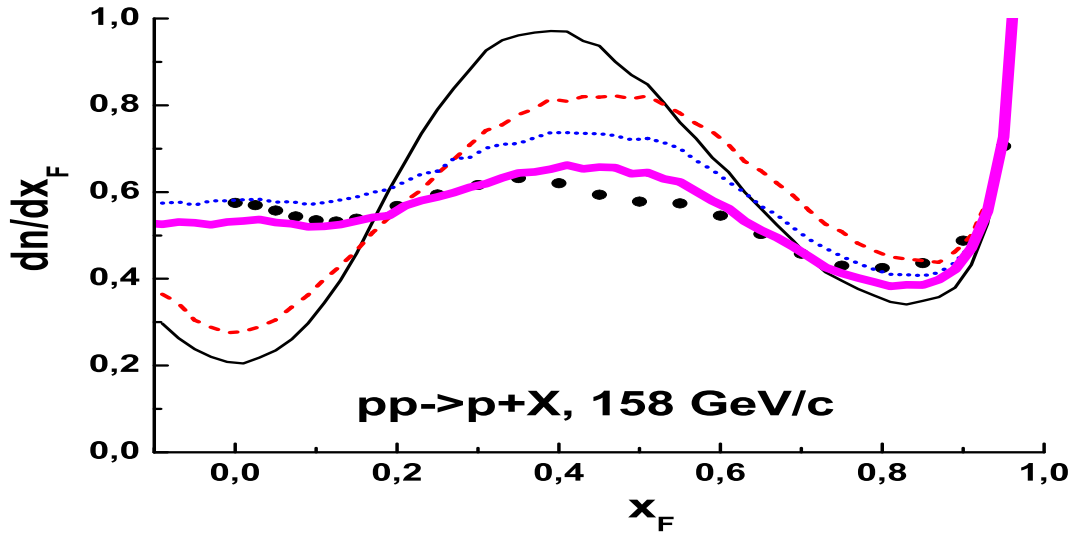}
\caption{$x_F$ distributions of protons in $pp$ interactions at 158 GeV/c. Closed points
are the NA49 experimental data \protect{\cite{5}}. Lines are results of FTF model simulations:
standard proton splitting (solid black), optimal diquark fragmentation function (dashed red),
string inversion (dotted blue), diquark suppression (1/6 instead of 1/3) including the optimal 
fragmentation function and the string inversion (solid thick).}
\label{Fig1}
\end{minipage}
\end{center}
\end{figure}
\vspace{-5mm}

The Fritiof (FTF) model of Geant4 \cite{10} (release geant4.10.0)  was used for the simulations. The model
considers a diquark as a unit, and assumes the standard splitting of nucleons. As seen,
the model overestimates the data at $x_F\sim$ 0.5, and underestimates the data in
the central region ($x_F \sim 0$). For reproduction of the data at $x_F\sim$1,
the high mass diffraction dissociation was simulated with a probability of
10 (mb)/$\sigma^{in}_{pp}$.

The form of the proton spectrum depends, first of all, on the diquark fragmentation function
into baryons, $f(z)$. The following fragmentation functions were tested: $1/z$, $1/\sqrt{z}$,
$1$, $\sqrt{z}$, $z$, $z^2$. The most acceptable results were obtained with
$f(z)\propto \sqrt{z}$. (see dashed red line in Fig.~1).

To improve the prediction in the central region, string inversion was introduced.
The Fritiof model assumes that diquarks are leading particles after the splitting.
It is not so in the quark-gluon-string model \cite{11,12}, where quarks can sometimes be
the leading particles. String inversion takes this into account by assuming that
a quark is the leading particle in 25\% of cases. This improved the situation in the central
region (see dotted blue line in Fig.~1). Now it is clearly seen that the calculations
overestimate the multiplicity of the protons.

Changing the weight of the (uu)--d configuration in protons from 1/3 to 1/6 and using the optimal 
fragmentation function and the string inversion  allow one to reach a semi-quantitative 
description of the proton spectrum (see solid thick line in Fig.~1). Of course, the description 
is not perfect but it can be improved in future by a fine tuning of the model parameters.

The proposed approach gives also a possibility to improve the description of neutron and $\Lambda$-hyperon
spectra (see Fig.~2).
\begin{figure}[cbth]
\begin{center}
\begin{minipage}{14cm}
\includegraphics[width=140mm,height=40mm,clip]{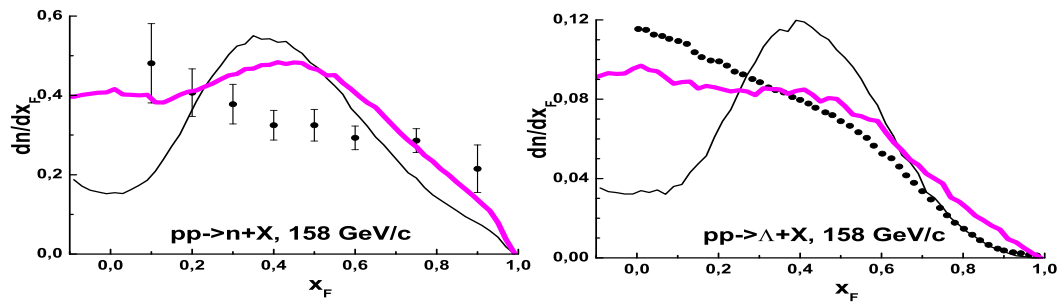}
\caption{$x_F$ distributions of neutrons and $\Lambda$-hyperon in $pp$ interactions at
158 GeV/c. Closed points are the NA49 experimental data \protect{\cite{5}}. Solid thin lines
are results of the standard FTF model simulations. Thick lines are results obtained by
suppression of the (uu)-diquarks.}
\label{Fig2}
\end{minipage}
\end{center}
\end{figure}
\vspace{-5mm}

In summary, we conclude that the weights of the (uu)--d configuration in protons and 
the (dd)--u configuration in neutrons have to be re-considered in Monte Carlo event generators.

The authors are thankful to N.~Kochelev for a useful consideration of the subject of the paper,
and J.~Harvey, D.H.~Wright, A.~Ribon for interest to the work and help.

\end{document}